\documentclass[
  dvipsnames,svgnames,x11names, 
  pmlr,                        
  anonymous,                   
  review,                      
  twocolumn,                   
  10pt                         
]{jmlr}

\usepackage{multirow}

\usepackage{placeins}

\usepackage[english]{babel}
\usepackage[utf8]{inputenc}
\usepackage[T1]{fontenc}
\usepackage{titletoc}
\usepackage{bbm}
\usepackage{bbold}

\usepackage[markup=highlight,inline]{changes}
\definechangesauthor[color=red]{qsX4}
\definechangesauthor[color=blue]{VLRR}
\definechangesauthor[color=ForestGreen]{tQ1u}
\definechangesauthor[color=yellow]{9snq}

\usepackage{hyperref}
\usepackage{url}
\usepackage{booktabs}
\usepackage{makecell}
\usepackage{doi}
\usepackage{adjustbox}
\usepackage{caption}

\usepackage{amsmath, amssymb, amsfonts, bbm}
\usepackage{nicefrac}
\usepackage{microtype}
\usepackage{cleveref}
\usepackage{booktabs}

\usepackage{graphicx}
\usepackage{float}
\usepackage{subcaption}
\usepackage{longtable}

\usepackage{tikz}
\usetikzlibrary{arrows.meta, positioning}
\usepackage{algorithm}
\usepackage{algpseudocode}
\usepackage{amsmath}

\usepackage{todonotes}
\setcommentmarkup{\todo[color={red!20},size=\scriptsize]{\centering \quad #1}}
\jmlrvolume{287}
\jmlryear{2025}
\jmlrsubmitted{} 
\jmlrpublished{} 
\jmlrworkshop{Conference on Health, Inference, and Learning (CHIL) 2025}

\usepackage[switch]{lineno}

\title[Semiparametric Regression for Treatment-Covariate Interactions]{Learning Interactions Between Continuous Treatments and Covariates  with a Semiparametric Model}

\author{%
\Name{Muyan Jiang} \Email{muyan\_jiang@berkeley.edu}\\
\addr Computational Precision Health, University of California, Berkeley, USA\\
\addr Department of Industrial Engineering and Operations Research, University of California, Berkeley, USA
\AND
\Name{Yunkai Zhang}\Email{yunkai\_zhang@berkeley.edu}\\
\addr Department of Industrial Engineering and Operations Research, University of California, Berkeley, USA\\
\AND
\Name{Anil Aswani} \Email{aaswani@berkeley.edu}\\
\addr Computational Precision Health, University of California, Berkeley, USA\\
\addr Department of Industrial Engineering and Operations Research, University of California, Berkeley, USA
}

\begin{document}

\maketitle

\begin{abstract}
Estimating the impact of continuous treatment variables (e.g., dosage amount) on binary outcomes presents significant challenges in modeling and estimation because many existing approaches make strong assumptions that do not hold for certain continuous treatment variables. For instance, traditional logistic regression makes strong linearity assumptions that do not hold for continuous treatment variables like time of initiation. In this work, we propose a semiparametric regression framework that decomposes effects into two interpretable components: a prognostic score that captures baseline outcome risk based on a combination of clinical, genetic, and sociodemographic features, and a treatment-interaction score that flexibly models the optimal treatment level via a nonparametric link function. By connecting these two parametric scores with Nadaraya-Watson regression, our approach is both interpretable and flexible. The potential of our approach is demonstrated through numerical simulations that show empirical estimation convergence. We conclude by applying our approach to a real-world case study using the International Warfarin Pharmacogenomics Consortium (IWPC) dataset to show our approach's clinical utility by deriving personalized warfarin dosing recommendations that integrate both genetic and clinical data, providing insights towards enhancing patient safety and therapeutic efficacy in anticoagulation therapy.
\end{abstract}

\paragraph*{Data and Code Availability}  
This paper uses public data from the International Warfarin Pharmacogenomics Consortium (IWPC) on the \href{https://www.pharmgkb.org/downloads}{Pharmacogenomics Knowledgebase}  \citep{whirl2021evidence}.  
Our code is available: \href{https://github.com/JimmyJiang666/Semiparametric-Regression-Treatment}{https://github.com/JimmyJiang666/Semiparametric-Regression-Treatment}.

\paragraph*{Institutional Review Board (IRB)}
This research does not require IRB approval. 

\section{Introduction}
In health-related decision-making, continuous treatments—such as dosing levels or intervention timing—play a critical role. However, estimating their impact on binary outcomes (e.g., treatment success or failure) remains a challenge. Traditional methods, such as logistic regression, impose rigid linearity assumptions that frequently fail to capture the complex, nonlinear interactions between treatment and patient-specific covariates \citep{thomas2014systematic}. In contrast, while flexible machine learning models like neural networks can capture these intricate relationships, their inherent lack of interpretability renders them less suitable for clinical settings where transparency is essential for trust and actionable decision-making.

In the context of continuous treatments and binary outcomes, regression models incorporating continuous treatment variables provide valuable insights into treatment-response dynamics, which are essential for optimizing interventions. Traditional approaches, such as logistic regression \citep{hosmer2013applied}, treat continuous treatments as linear covariates, imposing strong linearity assumptions that may fail to capture complex, nonlinear interactions between treatments and patient-specific covariates. More advanced techniques, such as outcome-weighted learning, have been developed to refine personalized dosing strategies \citep{chen2016personalized}. Similarly, \citet{kallus2018policy} introduced a framework for evaluating continuous treatment policies using kernel-enhanced inverse probability weighting and doubly robust methods, while \citet{jin2023change} proposed a change-plane model to identify and analyze subgroups with heterogeneous treatment effects. However, these methods often favor predictive accuracy over interpretability or trade transparency for flexibility, leading to complex models that hinder practical use in clinical settings.

Another related approach is the Generalized Propensity Score (GPS) method, which is useful for estimating causal effects of continuous treatments \citep{hirano2004propensity}. GPS methods rely on correctly specifying a parametric model for the treatment assignment given the covariates, and any misspecification can lead to biased estimates of the dose-response function. Although nonparametric or machine learning techniques can be employed to mitigate this issue, they add further complexity to the analysis.

Semiparametric regression models offer a promising alternative by integrating the strengths of both parametric and nonparametric approaches. A classical example is the single index model, given by
\[\mathbb{E}(Y\mid X) = g(\xi^T X) + \epsilon,\]
which has been successfully applied in fields ranging from economics to healthcare, providing a compact and interpretable summary of complex relationships \citep{hristache2001direct,d2022partially,wang2020family}. The term ``single index'' refers to capturing covariate effects via the univariate projection \( \xi^T X \), reducing dimensionality while maintaining flexibility through the nonparametric link function \( g \). Unlike GLMs where \( g \) is predefined (e.g., logit or identity), here it is estimated nonparametrically, often with smoothness constraints like monotonicity or differentiability. A spline-based \( g \), for instance, accommodates nonlinear relationships beyond the linear case \( g(u) = u \). Though traditionally applied to continuous \( Y \), extensions for binary outcomes make it a natural fit for our approach. Recent work has further extended these ideas to optimize treatment rules in clinical trials \citep{park2021constrained,park2022development}.

\begin{figure}[htbp]
\centering
\captionsetup{justification=justified,format=plain,indention=0pt} 
\includegraphics[width=1.0\linewidth]{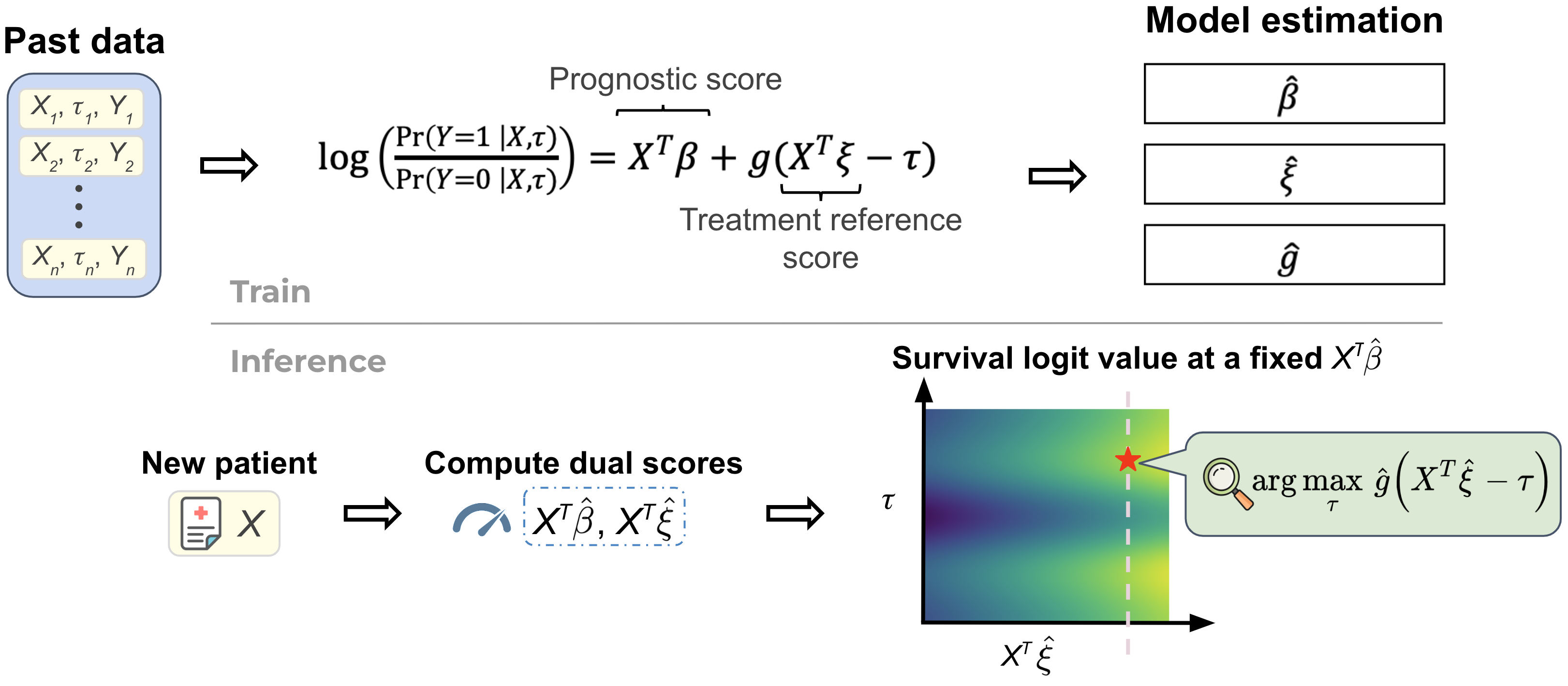}
\caption{Overview of our proposed pipeline. During training, the model learns $\hat{\beta}$ and $\hat{\xi}$ associated prognosis and treatment effectiveness, respectively. During inference, as the prognostic score $X^T\hat{\beta}$ is fixed conditioning on the patient's attributes, we can directly visualize outcome density using a heatmap with estimated $\hat{g}$. The optimal treatment value is then extrapolated at the maximum point along the y-axis corresponding to a fixed $X^T\hat{\xi}$.}
\label{fig:motivation}
\end{figure}

Building on these insights, our work proposes a semiparametric regression framework specifically tailored for settings with binary outcomes and continuous treatments. In our approach, the treatment effect is decomposed into two interpretable components: a \emph{prognostic score} that quantifies baseline risk and a \emph{treatment-interaction score} that captures how patient covariates modulate the treatment effect. The unknown treatment-interaction function is modeled nonparametrically via a flexible function, estimated using Nadaraya–Watson regression with regularization to ensure both robustness and interpretability. \figureref{fig:motivation} provides an overview of our pipeline.

The main contributions of this paper are as follows: 
\begin{enumerate}
\item A partially linear model for estimating individualized treatment strategies that maintain interpretability through quantifying effects.
\item A dual-score framework comprising (i) a prognostic score capturing baseline outcome risk and (ii) a treatment-interaction score, providing actionable guidance for treatment assignment.
\item Numerical simulations under diverse scenarios demonstrating empirical convergence and robustness of the estimation method.
\item A real-world application to anticoagulation therapy data, showcasing the model’s utility in deriving personalized warfarin dosing recommendations.
\end{enumerate}


\section{Model Setup and Estimation}

In clinical and observational studies, binary outcomes \( Y \in \{0,1\} \) (e.g., treatment success/failure) are often analyzed alongside covariates \( X \in \mathbb{R}^p \) and a continuous treatment variable \( \tau \in \mathbb{R} \). While standard logistic regression treats \(\tau\) as an additive covariate\footnote{Note that our model is general and allows both positive and negative treatment values, although medical treatments are typically positive (e.g., drug dosage, radiation intensity). However, some cases, such as temperature modulation or ventilator pressure adjustments, may involve both signs.}, this approach fails to address confounding---treatment assignment is often correlated with covariates, complicating effect estimation. To address this, we propose a model that explicitly incorporates treatment-covariate interactions while preserving interpretability through two clinically meaningful scores: a \textit{prognostic score} capturing baseline risk using covariates (analogous to the linear predictor in logistic regression) and a \textit{treatment-interaction score} guiding optimal \(\tau\) assignment based on covariate-driven interactions.

\subsection{Model Formulation}
Given binary response $Y\in\{0,1\}$, patient traits $X\in\mathbb{R}^p$, and treatment variable $\tau\in\mathbb{R}$, we consider the following logistic regression model:
\begin{equation}
    Pr(Y=1|X,\tau) = \sigma(X^T\beta + g(X^T\xi - \tau))
\end{equation}
where $\sigma(\cdot) = \frac{1}{1+exp(- \cdot)}$ is the logistic function, $g(\cdot):\mathbb{R}\to\mathbb{R}$ is an unspecified function that models the effect of the treatment, and $\beta, \xi\in\mathbb{R}^p$ are learnable parameters. The first prognostic score $X^T\beta$ models the main linear effect based on the baseline traits of a patient. The second treatment-interaction score $X^T\xi$ is related to the treatment effect, where the difference term $X^T\xi - \tau$ is the argument of an unknown function $g$ that encapsulates the covariates-treatment interaction. For instance, if \( g(u) = u^2 \), then \( g(X^T\xi - \tau) = (X^T\xi - \tau)^2 \), which introduces a multiplicative interaction between \( \tau \) and \( X^T\xi \), allowing the treatment effect to vary nonlinearly with patient covariates. Intuitively, $X^T\xi$ represents a patient-specific optimal treatment level derived from their covariates, and subtracting the actual treatment $\tau$ quantifies the deviation from this optimal level, which can be flexibly modeled by $g$. Furthermore, as we will discuss later, we normalize via $\xi \in \{\xi\in\mathbb{R}^p: \lVert \xi \rVert_2 = 1, \xi_1 \geq 0\}$ for identifiability because the function $g$ is unknown and can absorb the scale and sign of $\xi$. Function $g$ also needs to be learned alongside $\beta$ and $\xi$.\footnote{Specifically, since $g$ is estimated nonparametrically, any rescaling or sign reversal of $\xi$ can be offset by a corresponding transformation of $g$, making $\xi$ identifiable only up to scale and sign.}

We consider the ``odds ratio'' (OR), a measure of association between a treatment and an outcome, and obtain the following:
\begin{equation}\label{eq:pl}
    \log{\frac{Pr(Y=1|X,\tau)}{Pr(Y=0|X,\tau)}} = X^T\beta + g(X^T\xi - \tau),
\end{equation}
which reduces to a partially linear model with an unknown nonlinear part.

In a practical application, the prognostic score $X^T\beta$ highlights the pre-treatment condition, and the subsequent treatment-interaction score $X^T\xi$ is analyzed in conjunction with the function $g$, which elucidates the additive impact of the proposed treatment.
Given data $\{X_i,Y_i,\tau_i\}_{i=1,...,n}$, denote $\bar{Y}_i = \log{\frac{Pr(Y_i=1|X_i,\tau_i)}{Pr(Y_i=0|X_i,\tau_i)}}$, $Z_i(\xi) = X_i^T\xi - \tau_i$ for a fixed $\xi$. Our partially linear model becomes
\begin{equation}\label{eq:pl-data}
    \bar{Y}_i = X_i^T\beta + g(Z_i(\xi)) + e_i,
\end{equation}
where $e_i$ is the error term with $\mathbb{E}(e_i|Z_i(\xi)) = 0$.

\subsection{Proposed Estimation}

In the above formulation, it remains to estimate parameters $\beta,\xi$ and the unknown function $g$, which we denote as $\hat{\beta},\hat{\xi}$, and $\hat{g}$ respectively.

Condition \equationref{eq:pl-data} on $Z_i(\xi)$, we obtain 
\begin{equation}\label{eq:condition}
    \mathbb{E}(\bar{Y}_i|Z_i(\xi)) = \mathbb{E}(X_i^T\beta|Z_i(\xi)) + g(Z_i(\xi)).
\end{equation}
Taking the difference of \equationref{eq:pl-data} and \equationref{eq:condition}, we obtain the following least-squares (LS) regression:
\begin{equation}\label{eq:eyi}
    e_{yi} = e_{xi}^T\beta + e_i
\end{equation}
where $e_{yi} =  \bar{Y}_i -  \mathbb{E}(\bar{Y}_i|Z_i(\xi))$ and $e_{xi} = X_i - \mathbb{E}(X_i|Z_i(\xi))$.

Based on \equationref{eq:eyi}, we can immediately compute an (infeasible) estimate of $\beta$ for LS regression
\begin{equation*}
    \Tilde{\beta}(\xi) = \left(\sum_i^n e_{xi}e_{xi}^T\right)^{-1}\sum_{i}^n e_{xi}e_{yi}^T.
\end{equation*}
Note that this estimate depends on a realized value of $\xi$ so we denote $\Tilde{\beta}(\xi)$ to make the dependency clear. However, this estimate is infeasible because the unknown conditional means must be estimated. We can estimate $e_{xi},e_{yi}$ using nonparametric kernel regression such as the Nadaraya-Watson (NW) estimator \citep{nadaraya1964estimating,watson1964smooth} and obtain:
\begin{equation}\label{eq:beta-hat}
    \bar{\beta}(\xi) = \left(\sum_i^n \hat{e}_{xi}\hat{e}_{xi}^T\right)^{-1}\sum_{i}^n \hat{e}_{xi}\hat{e}_{yi}^T,
\end{equation}
where 
\begin{equation}\label{eq:nw-x}
    \hat{e}_{xi} = X_i - \frac{\sum_{j\neq i}^n K_{h_i}(Z_j(\xi)-Z_i(\xi)) \cdot X_j}{\sum_{j\neq i}^n K_{h_i}(Z_j(\xi)-Z_i(\xi))},
\end{equation}
\begin{equation}\label{eq:nw-y}
    \hat{e}_{yi} = \bar{Y}_i - \frac{\sum_{j\neq i}^n K_{h_i}(Z_j(\xi)-Z_i(\xi)) \cdot \bar{Y}_j}{\sum_{j\neq i}^n K_{h_i}(Z_j(\xi)-Z_i(\xi))}.
\end{equation}

Here, \( K(\cdot): \mathbb{R}\to\mathbb{R}^{+} \) is any kernel function that satisfies \( \int K(t)dt=1 \), \( \int tK(t) dt = 0 \), and \( 0 < \int t^2K(t) dt < \infty \). Typical choices include the Epanechnikov kernel, \( K(t) = \frac{3}{4}(1-t^2)\,\mathbb{1}_{\{|t|\leq 1\}} \), the Gaussian kernel, \( K(t) = \frac{1}{\sqrt{2\pi}} e^{-t^2/2} \), and other commonly used smoothing functions. The latter terms in both \equationref{eq:nw-x} and \equationref{eq:nw-y} correspond to the Nadaraya–Watson (NW) estimators, where \( h_i \) is the selected bandwidth, which serves as a hyperparameter that controls the degree of smoothing.

Note that an estimate of $\hat{\beta}$ depends on a given $\xi$, so we first estimate $\hat{\xi}$ by minimizing estimated residuals as follows:
\begin{equation}\label{obj:original}
    \hat{\xi} = \operatorname*{argmin}_{\xi} \sum_i^n ( r_i(\xi))^2,
\end{equation}
where $r_i(\xi) = \hat{e}_{yi} - \hat{e}_{xi}^T\bar{\beta}(\xi)$.

We then use the obtained $\hat{\xi}$ to compute the corresponding $\hat{\beta} = \bar{\beta}(\hat{\xi})$ from \equationref{eq:beta-hat}. Finally, we can estimate $\hat{g}$ in the following nonparametric way using NW regression again:
\begin{equation}\label{eq:g-hat}
    \hat{g}(Z) = \frac{\sum_{i}^n K_{h_i}(Z_i(\hat{\xi})-Z) \cdot (\bar{Y}_i - X_i^T\hat{\beta})}{\sum_{i}^n K_{h_i}(Z_i(\hat{\xi})-Z)}.
\end{equation}

This approach is implicitly an iterative procedure for the parameter estimation. However, because of the closed-form OLS solution for $\bar{\beta}(\xi)$ within the optimization process for $\xi$, we can effectively perform a one-run optimization on $\xi$ first, rather than alternating between the two parameter estimates. We show its pseudocode in Algorithm~\ref{alg:estimation} in the Appendix.

In this model, the coefficient \( \xi \) is identifiable only up to scale and sign because the nonparametric function \( g \) can absorb rescalings and sign changes of \( \xi \), making \( \xi \) unique only up to these transformations. To ensure identifiability, we impose the constraint  
\[
\Theta = \{\xi = (\xi_1,...,\xi_p)^T \in \mathbb{R}^p | \left\| \xi \right\|_2 = 1 , \xi_1 \geq 0\}.
\]
In addition, LASSO regularization is incorporated to promote sparsity in the estimation of $\xi$ \citep{tibshirani1996regression,hoerl1970ridge}. This regularization is particularly important in healthcare applications where high-dimensional data are common. Sparsity not only helps mitigate issues of multicollinearity but also simplifies the model, allowing clinicians to easily identify the most influential patient characteristics. Therefore, on top of the constrained version of \equationref{obj:original}, we add a lasso penalty to $\xi$ and estimate the model by optimizing the following:
\begin{align}\label{obj:constrained}
    \hat{\xi} &= \operatorname{argmin}_{\xi} \sum_i^n ( r_i(\xi))^2 + \lambda \left\| \xi \right\|_1 \\
    & \text{s.t.} \quad \left\| \xi \right\|_2 = 1 , \xi_1 \geq 0. \notag
\end{align}
Notably, we regularize \( \xi \) but not \( \beta \) because \( \xi \) captures treatment-covariate interactions, which are often high-dimensional and complex. In contrast, \( \beta \) models baseline prognosis and is left unregularized to provide an unbiased estimate, consistent with traditional logistic regression. The bandwidth and lasso penalty $\lambda$ are hyperparameters tunable via cross-validation.

\begin{table*}[thpb]
\caption{Setup of simulated scenarios.}
\centering
\scriptsize 
\setlength{\tabcolsep}{4pt} 
\begin{tabular}{l|c|c|c|c}
\hline
 &\textbf{Covariates Type} & \textbf{Partially Linear Interaction} & 
 \makecell{\textbf{Interation Type}}& \textbf{Intervention Setting} \\ \hline
 Scenario 1& Continuous &  $\bar{Y} = X^T \beta + 3$ &  Constant &  RCTs\\ \hline
 Scenario 2& Continuous &  $\bar{Y} = X^T \beta + (X^T \xi - \tau)$&  Linear &  RCTs\\ \hline
 Scenario 3& Continuous &  $\bar{Y} = X^T \beta - 0.5 \log(|X^T \xi - \tau|)$ & Unimodal &  Observational\\ \hline
 Scenario 4& Categorical \& Continous &  $\bar{Y} = X^T \beta -1.2 cos(\pi \cdot (X^T \xi - \tau)) \cdot e^{-(X^T \xi - \tau)^2} $ &  Multimodal&  RCTs\\ \hline
\end{tabular}
\label{table:table_all}
\end{table*}

\section{Numerical Simulations for Empirical Convergence Analysis}
In our simulation study, we designed four scenarios (summarized in \Cref{table:table_all}) intended to systematically evaluate the convergence of the proposed estimation framework in different settings commonly encountered in clinical studies. Scenarios 1 and 2 represent randomized control trial (RCT)-like settings with simple interaction patterns (constant and linear, respectively) serving as baseline performance benchmarks. Scenario 3 simulates observational data to evaluate robustness against confounding, with a more complex unimodal interaction. Scenario 4 assesses the method's ability to handle high-dimensional mixed data (categorical and continuous) and complex multimodal interactions.

\subsection{Experimental Setup}
For scenario 1 and 2, $X$ are generated as a multivariate Gaussian distribution with random mean $\mu$ and covariance matrix $\Sigma$ where $\mu \sim Unif(-1,1)^p$ and $\Sigma = AA^T$ with $A \sim Unif(0,1)^{p\times p}$. The treatment variable is generated using a standard Gaussian distribution $\tau \sim \mathcal{N}(0,1)$. For the parameters, we simulate $\beta \sim \text{Unif}(-1, 1)^4$ and $\xi \sim \text{Unif}(\mathbb{Q}_1)$ where $\mathbb{Q}_1$ denotes the first quadrant on the unit sphere due to the identifiability constraint. The treatment-covariate term is constant in scenario 1 and linear in scenario 2. For the constant term in scenario 1, it means treatment assignment does not affect the outcome. The linear term in scenario 2 implies that the treatment acts linearly on the outcome, and the optimal treatment decision is always the extreme point within the feasible range of the treatment, depending on the coefficient sign of the variable. In both scenarios, we let $p = 8$.
    
\begin{figure}[htbp]
\centering
\begin{tikzpicture}[node distance=0.5cm and 0.5cm, scale=0.85, transform shape] 
    \tikzset{
      >={Latex[width=1mm,length=1mm]}, 
      base/.style = {circle, draw=black,
                     minimum width=1em, text centered, 
                     font=\sffamily},
      X1/.style = {base},
      X2/.style = {base},
      X3/.style = {base},
      X4/.style = {base},
      Tau/.style = {base}
    }

    \node[X1] (x1) {$X_1$};
    \node[X2] (x2) [right=of x1] {$X_2$};
    \node[X3] (x3) [below right=of x1] {$X_3$};
    \node[X4] (x4) [left  = of x3] {$X_4$};
    \node[Tau] (tau) [below right=of x2] {$\tau$};

    \path[->] (x1) edge (x2);
    \path[->] (x1) edge (x3);
    \path[->] (x2) edge (x3);
    \path[->] (x4) edge (x1);
    \path[->] (x2) edge (tau);
    \path[->] (x3) edge (tau);
\end{tikzpicture}
\caption{Directed Acyclic Graph in scenario 3.}
\label{fig:dag1}
\end{figure}
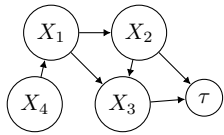

For scenario 3, we mimic an observational setting in which the treatment variable is intrinsically influenced by the covariates, as illustrated in \figureref{fig:dag1}. We simulate $X_4$ as an independent uniform random variable from -1 to 1: $X_4 \sim \text{Unif}(-1, 1)$. Then we simulate $X_1 = \sqrt{|X_4|} + \text{Unif}(-1, 1)$, $X_2 = 0.5 \times X_1 + \text{Unif}(-0.5, 0.5)$, $X_3 = 0.3 \times X_1 + 0.3 \times X_2 + \text{Unif}(-0.4, 0.4)$ which are correlated. For the treatment, we have $\tau = \sin(X_2X_3) + \text{Unif}(-0.6, 0.6)$, which are influenced by covariates $X_2$ and $X_3$. Similarly, we simulate $\beta \sim \text{Unif}(-1, 1)^4$ and $\xi \sim \text{Unif}(\mathbb{Q}_1)$. Lastly, for the interaction term, we have $\bar{Y} = X^T \beta - 0.5 \log(|X^T \xi - \tau|)$, where the optimal treatment is an unimodal function in the logarithmic term. 

Notably, scenario 3 differs from the others in that \( \tau \) explicitly depends on \( X_2 \) and \( X_3 \), as illustrated in a causal Directed Acyclic Graph (DAG) \figureref{fig:dag1}. This setup mirrors an observational study where treatment assignment is influenced by patient characteristics. In contrast, in all other scenarios, \( \tau \) is independent of the covariates, resembling an RCT. Moreover, the treatment-covariates interaction function in scenario 3 assumes a unique optimal treatment captured by the log function with the argument in absolute value. The assumption is that for every patient, given the covariates, there exists a unique optimal treatment that maximizes the outcome variable.

\begin{figure*}[t]
\centering
\begin{adjustbox}{max width=\textwidth,center}
\includegraphics[width=1.0\textwidth]{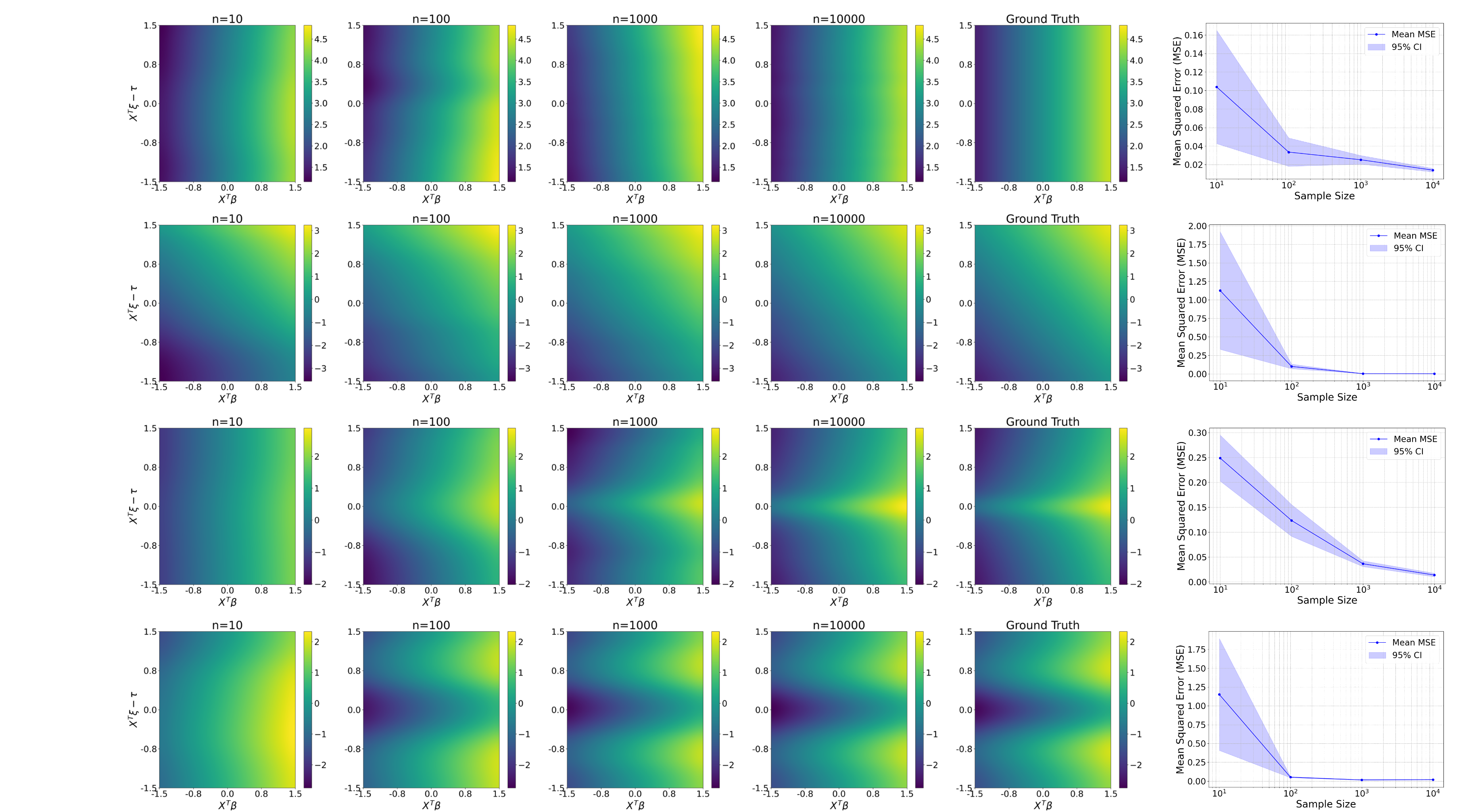}
\end{adjustbox}
\caption{Empirical Convergence of Estimations. The four rows represent scenario 1-4, respectively. The heatmaps show estimated values of the modeled function across two dimensions: the prognostic score ($X^T\beta$, x-axis) and the covariate-treatment term ($X^T\xi - \tau$, y-axis). The colored density of the heatmap represents $\bar{Y} = \log{\frac{Pr(Y=1|X,\tau)}{Pr(Y=0|X,\tau)}}$, the log-odds ratio, where higher values indicate a greater likelihood of the positive outcome. Columns in the heatmap panel represent increasing sample sizes ($n = 10$ to $10^4$), illustrating progressive convergence towards the true underlying function, shown in the rightmost (Ground Truth) column. The far-right plots quantify convergence through Mean Squared Error (MSE), bootstrapped ten times for confidence intervals. Note that the scale of the MSE y-axis differs between scenarios for visual clarity.}
\label{fig:experiement_all}
\end{figure*}

For scenario 4, we want to simulate our covariates as a mixture of categorical and continuous variables in an RCT setting where the treatment variable is independent of the covariates. Let $n_c,n_b$ denote the number of continuous and binary variables. We simulate $X_c \sim \mathcal{N}(\mu_c,\Sigma_c)$ with $\mu_c \in \mathbb{R}^{n_c}, \Sigma_c \in \mathbb{S}^{n_c}_{+}$, similarly as in scenario 1 and 2. For binary variables, we simulate $X_{b}[1] \sim \text{Bernoulli}(p_1)$ with $p_1 \sim Unif(0,1)$. For $i = 2, ..., n_b$, we simulate $X_b[i] = \text{Bernoulli}(p_i) $ with $p_i = \frac{1}{1 + \exp(-(a_i \cdot X_b[i-1] + b_i))}$. For each subsequent binary variable $X_b[i]$,  its Bernoulli probability $p_i$ is determined using a logistic function applied to a linear combination of the previous binary variable $X_b[i-1]$. The coefficients $a_i$ and $b_i$ in the logistic function can be adjusted to control the strength of the dependency. Notice that the treatment is generated independent of the covariates to resemble an RCT scenario. In our generation, we set $n_c=12, n_b = 8, a_i = 0.5, b_i = -0.25$, to imitate realistic clinical datasets where continuous covariates typically outnumber categorical ones, reflecting more extensive clinical measurements compared to binary indicators (e.g., conditions or genotypes). The parameters $a_i$ and $b_i$ were chosen to induce moderate dependencies among categorical variables, realistically representing clinically plausible correlation structures without creating overly strong collinearity. We generate $\tau \sim Unif(-1,1)$ independent of the covariates for the treatment. The treatment-covariates interaction term in scenario 4 involves some trigonometric and exponential functions. This creates some multimodal locally optimal treatment decisions, which can be seen in the heatmap of scenario 4 in \figureref{fig:experiement_all}. 

\tableref{table:table_all} summarizes the design for all scenarios. We estimated the model for each scenario using the constrained optimization framework as delineated in \equationref{obj:constrained}. We investigated and implemented a range of optimization strategies, including classical methods such as Differential Evolution \citep{storn1997differential} and CMA-ES \citep{hansen2001completely}. After assessing their performance in terms of computational efficiency, convergence behavior, and the ability to explore constrained search spaces, we ultimately implemented the Tree-based Parzen Estimators (TPE) algorithm via the \textit{hyperopt} library \citep{bergstra_hyperopt-proc-scipy-2013}. This probabilistic optimization framework, as described in \citep{watanabe2023tree}, is well-suited for this problem as it adaptively models the objective function's landscape using a probabilistic approach, efficiently focusing on high-promise regions while respecting the defined constraints on \( \xi \in \Theta \). Unlike unconstrained algorithms that may propose solutions outside the feasible space or inefficiently explore irrelevant regions, TPE ensures that estimates remain within the valid domain.

For clarity, the simulations assess empirical convergence to the ground-truth function rather than predictive accuracy via train-test splits. Each scenario generates independent samples (\( n = 10 \) to \( 10^4 \)) to evaluate estimation consistency. Hyperparameters (\(\lambda, h\)) are selected via cross-validation at each sample size. Since the focus is on convergence, all generated samples are used for estimation, and accuracy is measured using the mean squared error (MSE), defined as \( \int (\hat{g}(Z) - g(Z))^2 dZ \). This integral quantifies the deviation between the estimated and true function over the entire domain, making it a natural metric for assessing convergence. The far right column in \figureref{fig:experiement_all} reports MSE trends with bootstrapped confidence intervals.

\subsection{Results and Discusion}
In scenario 1, characterized by a constant interaction term, convergence towards the ground truth is achieved swiftly, even with a limited number of training samples. Notably, with $n = 10000$, the estimated heatmap accurately reflects the treatment's non-influence, maintaining a consistent scale along the y-axis. A similar result is observed in scenario 2, where the interaction term follows a linear pattern. This linear relationship manifests as a diagonal gradient in the heatmap, where the effect of treatment changes proportionally with the covariate-treatment interaction score. Scenario 3, featuring an unimodal interaction term, requires a larger sample size to align with the ground truth. This unimodal effect is visible in the heatmap as a single yellow band, indicating the unique optimal treatment level for different patient covariates. Starting from $n=1000$, the estimation progressively discerns the unique optimal treatment, as captured by the logarithmic function. Scenario 4 introduces increased dimensionality, more complex data types, and a multimodal interaction term. The presence of multiple optimal treatment levels is reflected in the heatmap by two distinct yellow bands, illustrating the multimodal nature of the treatment response. The estimation necessitates a larger number of samples to achieve empirical convergence and discern multimodality in optimal treatment decisions.

For all scenarios, MSE decreases consistently as the sample size increases. Scenarios 1–3 exhibit notably lower MSE compared to Scenario 4, reflecting their simpler interaction structures and lower dimensionality. Importantly, our estimation framework demonstrates empirical convergence across diverse data types (continuous and binary covariates) and settings (RCTs and observational). Although observational data typically pose challenges due to confounding, which can reduce accuracy, we observe successful convergence in our simulations. Note that our observational scenario employs a simplified confounding structure; thus, further study with more complex, realistic confounding is needed to fully assess the method’s robustness in practical settings.

\section{Application to Personalized Warfarin Dosing}
We evaluate our semiparametric regression framework on a real-world dataset from the International Warfarin Pharmacogenomics Consortium (IWPC) \citep{whirl2021evidence} to provide insights on individualized warfarin dosing strategies that enhance therapeutic efficacy, by integrating diverse clinical and genetic data. This dataset consists of 6000 patients. Warfarin is a critical anticoagulant with a narrow therapeutic range, where personalized dosing is essential due to marked inter-individual variability in drug response \citep{kim2015simplified}. The challenge of optimal dosing is particularly critical in intensive care and other high-risk settings, where rapid and precise dose adjustments are essential for patient safety.

\begin{figure}[htbp]
    \centering
\includegraphics[width=0.75\linewidth]{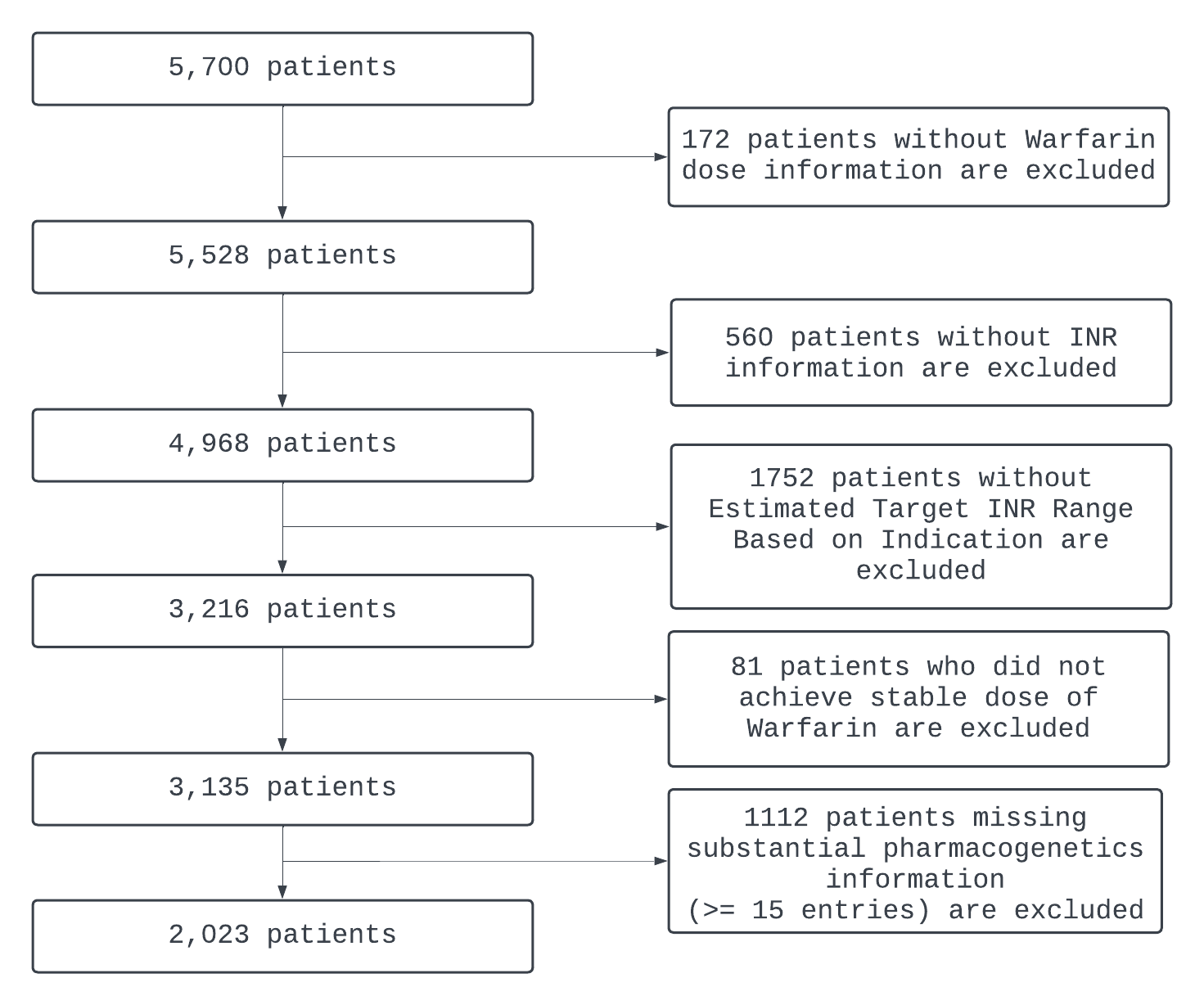}
    \caption{Cohort selection in the IPWC dataset.}
    \label{fig:cohortselection}
\end{figure}

The cohort selection process is shown in \figureref{fig:cohortselection}. The selected cohort is divided into train and test sets with a ratio of 9:1. We extracted 30 covariates with both clinical and pharmacogenetic variables, including physical attributes, medical conditions, concomitant medications, genotype status of functional warfarin genetic polymorphisms, etc. Variables were selected based on their clinical relevance to warfarin metabolism, as informed by prior pharmacogenetic studies \citep{dietz2020pharmacogenetic,wang2021warfarin,kim2015simplified}, as well as non-missingness in the dataset. Additionally, feature importance analysis from the soft label distillation process (see Section \ref{sec: soft-label}) also guided the inclusion of predictive covariates. In particular, we included CYP2C9 and VKORC1 polymorphisms due to their well-documented roles in warfarin metabolism. CYP2C9 encodes a cytochrome P450 enzyme that metabolizes warfarin, with variants (*2, *3) reducing enzymatic activity, slowing clearance, and increasing bleeding risk. VKORC1 encodes warfarin’s target, vitamin K epoxide reductase, with variants (497, 1173, 1542, 1639, 3730) affecting enzyme expression and dose requirements. These markers were selected based on their established clinical relevance in pharmacogenetic studies. In practice, pharmacogenetic-guided dosing can improve dosing effectiveness in highly sensitive responders versus that of patients who received fixed-dose \citep{dietz2020pharmacogenetic}. The International Normalized Ratio (INR) is a standardized measure of blood coagulation used to monitor the effectiveness of warfarin therapy. It represents the ratio of a patient’s prothrombin time to a control sample. Maintaining an INR within the appropriate target range is crucial to balancing the risk of clotting and bleeding. Different from studies that considered a fixed target INR \citep{park2022single}, we constructed a binary outcome that indicates whether a patient's INR falls within their personalized target range. This individualization is important since for deep vein thrombosis and atrial fibrillation, the target is 2.0–3.0; for high-risk heart valve patients, it's above 3.0; a 1.5–2.0 range is recommended for some heart valve patients to reduce bleeding risks \citep{puskas2018anticoagulation}.

In this study, our objective \textit{diverges from} intensively refining the accuracy of warfarin dosage predictions, as explored in previous research \citep{chen2023algorithmic,wang2021warfarin,lee2021development}. Instead, our model focuses on the ability to simultaneously quantify the patient’s baseline risk via the prognostic score and delineate the optimal treatment level through the treatment-interaction score.

\subsection{Distillation of Soft Labels} \label{sec: soft-label}

Because only binary labels are available for training, this causes the left-hand side of \equationref{eq:pl-data}, $\bar{Y}_i = \log{\frac{Pr(Y_i=1|X_i,\tau_i)}{Pr(Y_i=0|X_i,\tau_i)}}$, to be ill-defined. To address this issue, we train an intermediary model and then use the prediction of each sample (i.e., soft labels) in lieu of the binary ground truth to construct $\bar{Y}_i$. We refer to this intermediary model as the ``expert model'' and our model as the ``student model'' following \citet{hinton2015distilling}. In our implementation, we used XGBoost \citep{chen2016xgboost} as the expert model.

\begin{figure}[htbp]
    \centering
\includegraphics[width=0.65\linewidth]{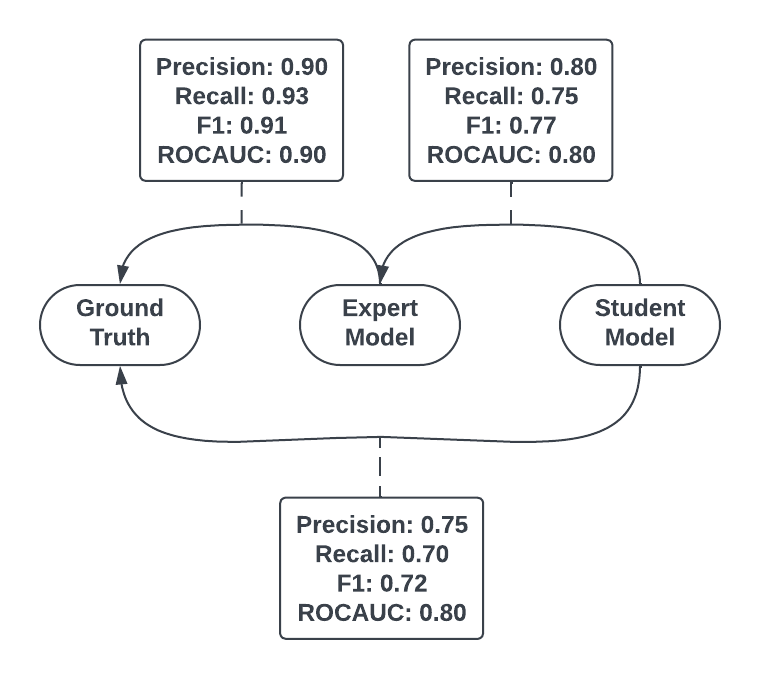}
    \caption{Predictive performance comparison. The arrow indicates the comparison's truth benchmark.}
    \label{fig:pred_perf}
\end{figure}

For preprocessing, we normalized continuous variables and one-hot encoded categorical ones. Due to label imbalance (negative samples ~12\%), we applied SMOTE \citep{chawla2002smote} to achieve a 1:1 ratio.
The expert model’s performance is shown in the top-left panel of \figureref{fig:pred_perf}. The expert model assigns soft labels used to train the student model.

\subsection{Student Model Training}

We aligned the train, validation, and test split with the expert model and estimated the student model. To reduce computational load, we chose the Epanechnikov kernel \( K(t) = \frac{3}{4}(1 - t^2)\mathbb{1}_{\{|t| \leq 1\}} \) for its finite support, limiting computations to local neighborhoods. In practice, kernel choice had minimal impact on results, provided a reasonable bandwidth was used.

We tuned the two key hyperparameters in our model—the kernel bandwidth \( h \) and the LASSO penalty parameter \( \lambda \)—using 5-fold cross-validation, selecting values that minimized the average mean squared error (MSE). See Appendix \ref{sec: hyperparameter-sensitivity} for the sensitivity analysis on these two hyperparameters.

As a classification model, the student model's performance is constrained by the expert model because the student model does not have direct access to the ground truth and is only trained by the expert model, as shown in \figureref{fig:pred_perf}. Using a more sophisticated and well-specified expert model may improve the accuracy. 

\subsection{Comparison on Other Prognostic Scores}\label{sec:comp_diag_score}

In our model,  $X^T\beta$ provides a prognostic score of the patient. This term can be analogized to the linear predictor in logistic regression. We compare the prognostic score estimation between our model and other linear models that offer similar insights, including Logistic Regression (LR) and  Linear Discriminant Analysis (LDA), each with a variant in which the treatment variable is incorporated as a continuous covariate. 

\begin{figure}[t]
\includegraphics[width=1.0\linewidth]{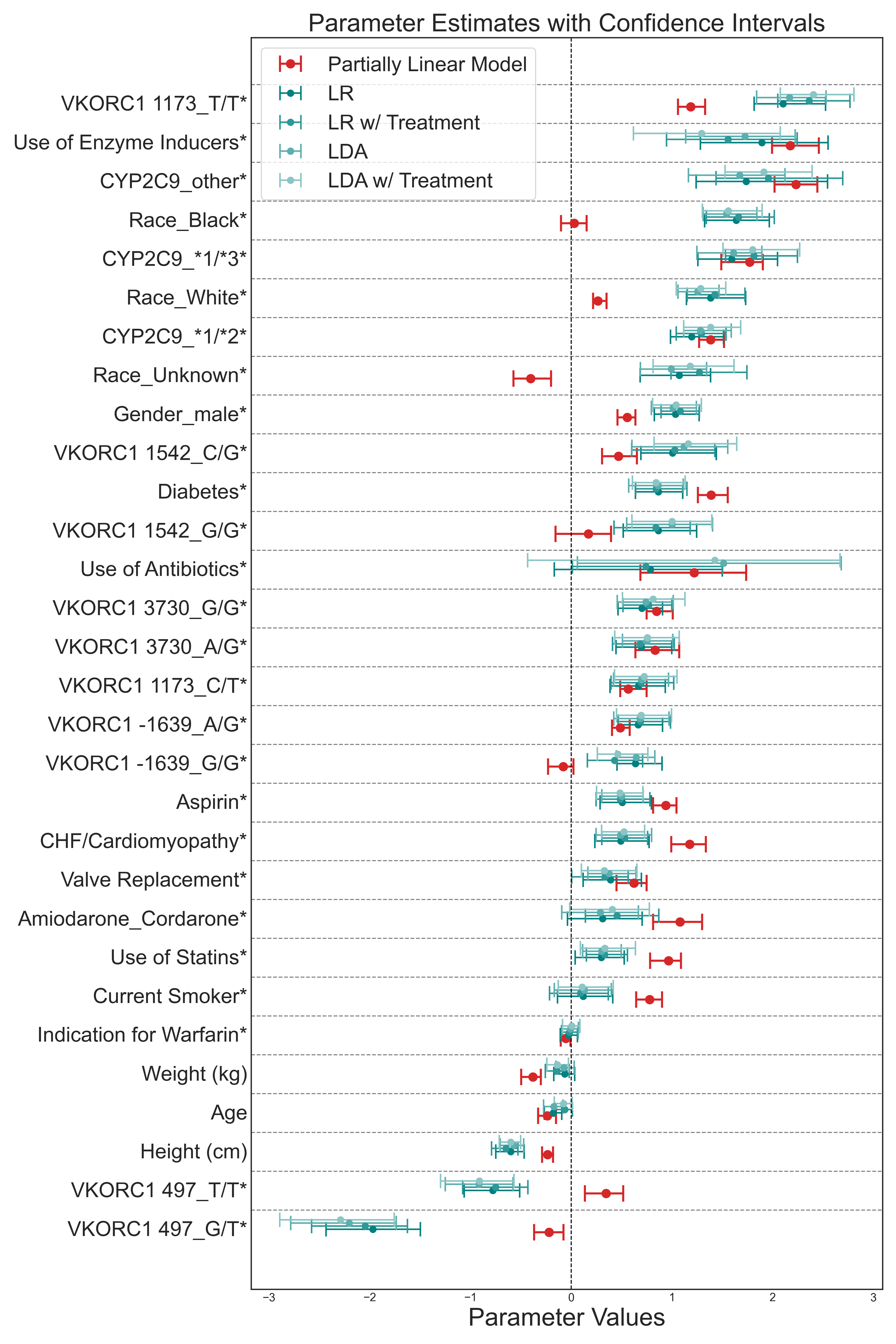}
\caption{Linear prognostic score comparison. Attributes are ordered by the magnitude of the LR coefficient estimates. Categorical features are marked with an asterisk sign.}
\label{fig:diag_score_comparison}
\end{figure}

The comparison in \figureref{fig:diag_score_comparison} demonstrates that, while there is general agreement in the direction of coefficients across models, our model exhibits notable differences in magnitude and confidence intervals for certain features. Specifically, our model moderates the influence of genetic polymorphisms, often shrinking their estimated effect toward zero, while assigning greater weight to clinical variables such as height, weight, and comorbidities (e.g., diabetes, use of statins). This suggests that our method prioritizes clinically relevant factors over genetic markers, which may have important implications for personalized treatment strategies.

Traditional models like LR and LDA assume a simple additive relationship between features and outcomes, while our approach explicitly incorporates nonlinear treatment-covariate interactions. Consequently, these differences lead to variations in coefficient estimates and confidence intervals. This reflects different modeling choices rather than limitations of any single method.

While these models are fundamentally different in structure, we compare them not to establish correctness but to illustrate how different modeling choices influence prognostic scores derived from quantifiable coefficients. Since no single model serves as an absolute benchmark for correctness, our goal is to open up the discussion on how different modeling frameworks yield different prognostic estimates, each with its own interpretability and clinical implications. Future work will involve consulting with domain experts to assess the clinical plausibility of our findings and validate the model’s utility in real-world treatment decisions.

\subsection{Clinical Interpretations of Estimates}

\begin{figure}[htb]
    \centering
\includegraphics[width=1.0\linewidth]{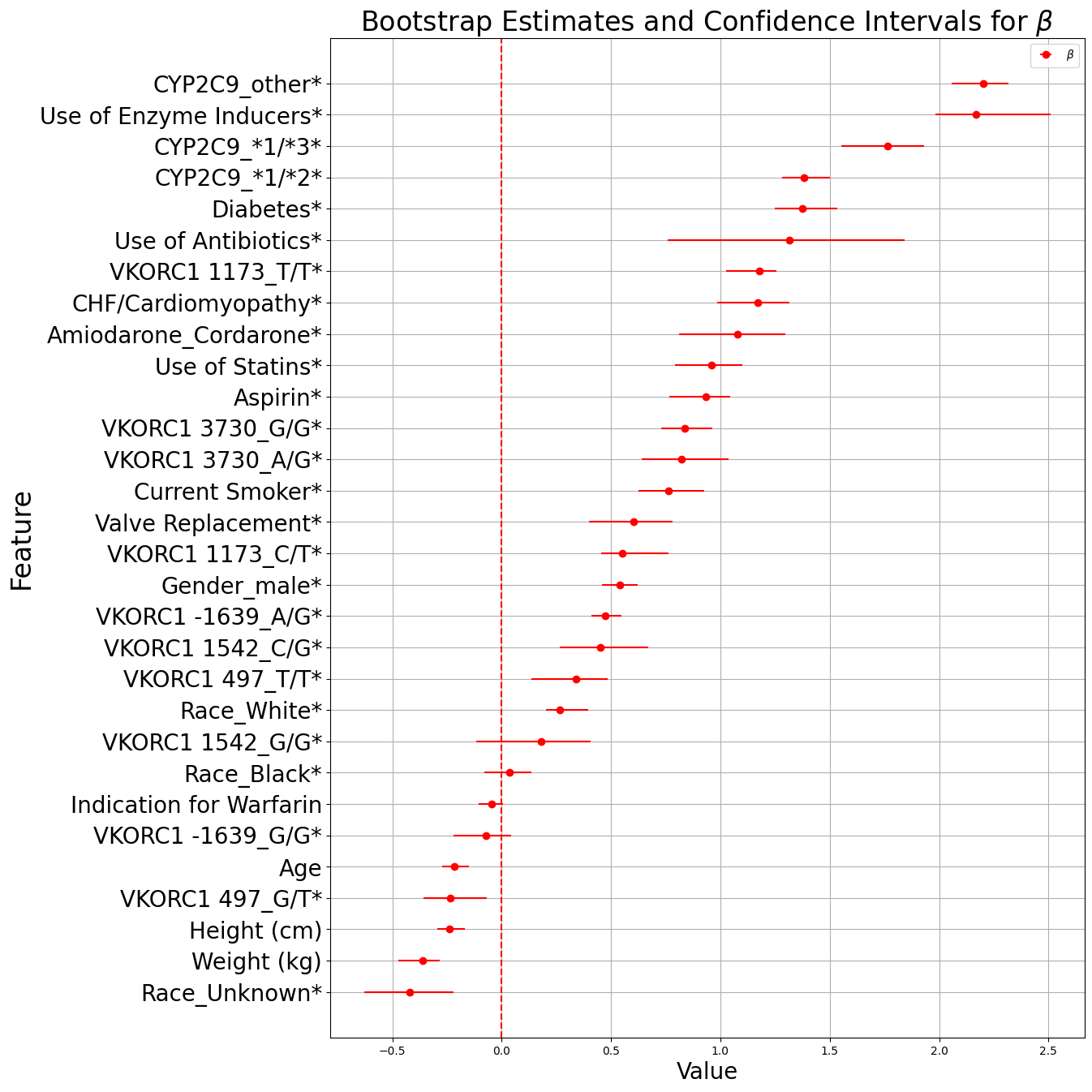}
    \caption{Values of $\beta$. Attributes are ordered by the magnitude of the corresponding coefficient estimates. Categorical features are marked with an asterisk sign.}
    \label{fig:ipwc_beta_conf}
\end{figure}

\begin{figure}[htbp]
    \centering
\includegraphics[width=1.0\linewidth]{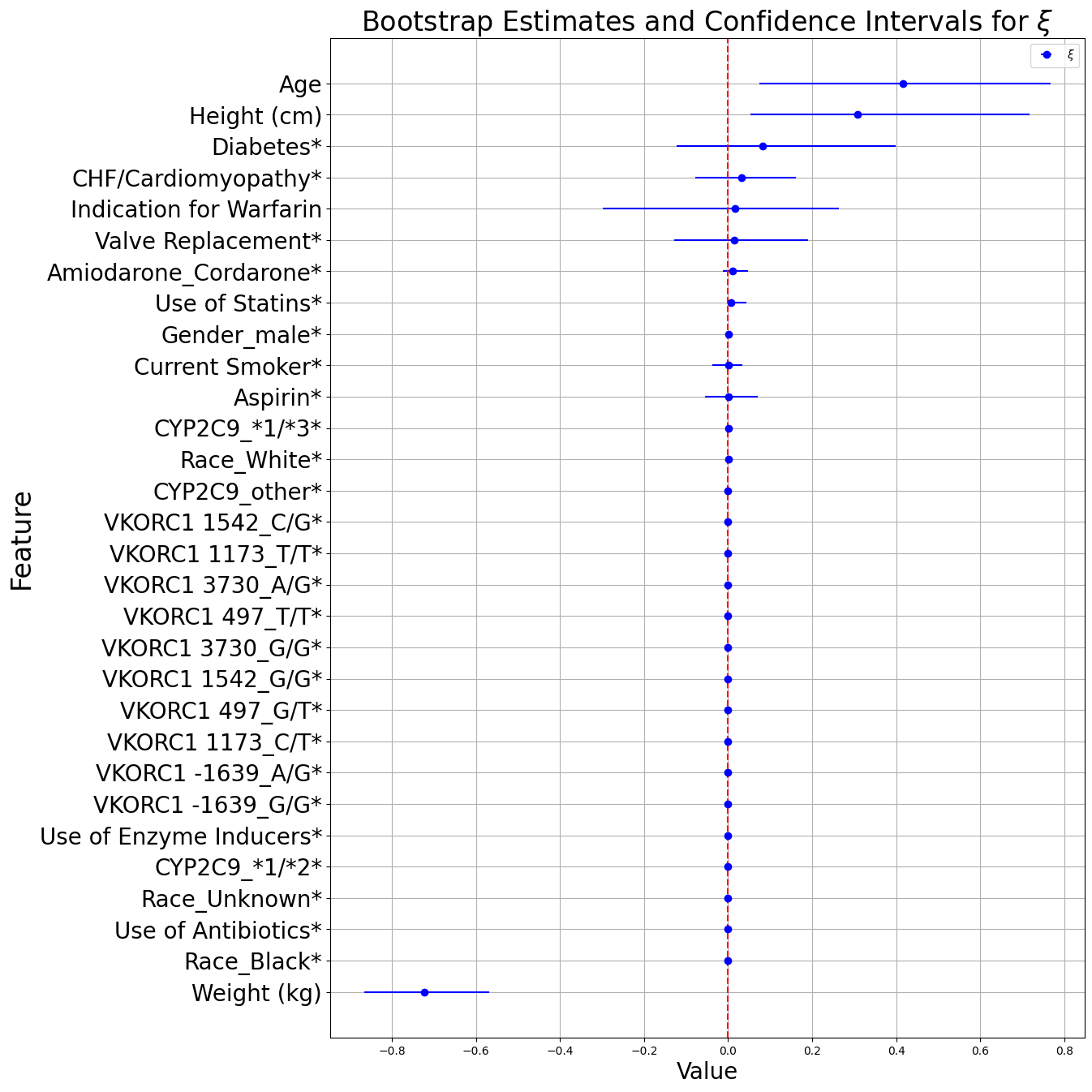}
    \caption{Estimated values of $\xi$. Attributes are ordered by the magnitude of their corresponding coefficient estimates. Categorical features are marked with an asterisk sign. A LASSO penalty is applied to $\xi$. Here, regularization eliminated over half the variables, highlighting the most relevant covariates for treatment effect modulation.}
    \label{fig:ipwc_xi_conf}
\end{figure}

\figureref{fig:ipwc_beta_conf} and \ref{fig:ipwc_xi_conf} show the estimated coefficients with bootstrapped confidence intervals with $k = 30$ iterations.

For $X^T\beta$, \figureref{fig:ipwc_beta_conf} indicates that larger values for ``age'', ``height'', and ``weight'' have a negative association with pre-treatment and may adversely affect the patient's ability to achieve the target INR range. This effect may be attributed to higher values for these variables generally corresponding to larger body sizes. Additionally, specific genotypes, such as variants of CYP2C9, are linked to diminished enzyme activity. Individuals carrying these alleles typically metabolize warfarin more slowly, thereby facing a higher risk of bleeding at conventional doses. Consequently, they often necessitate a lower warfarin dose to attain the target range.

\begin{figure}[htbp]
    \centering
    \includegraphics[width=\linewidth]{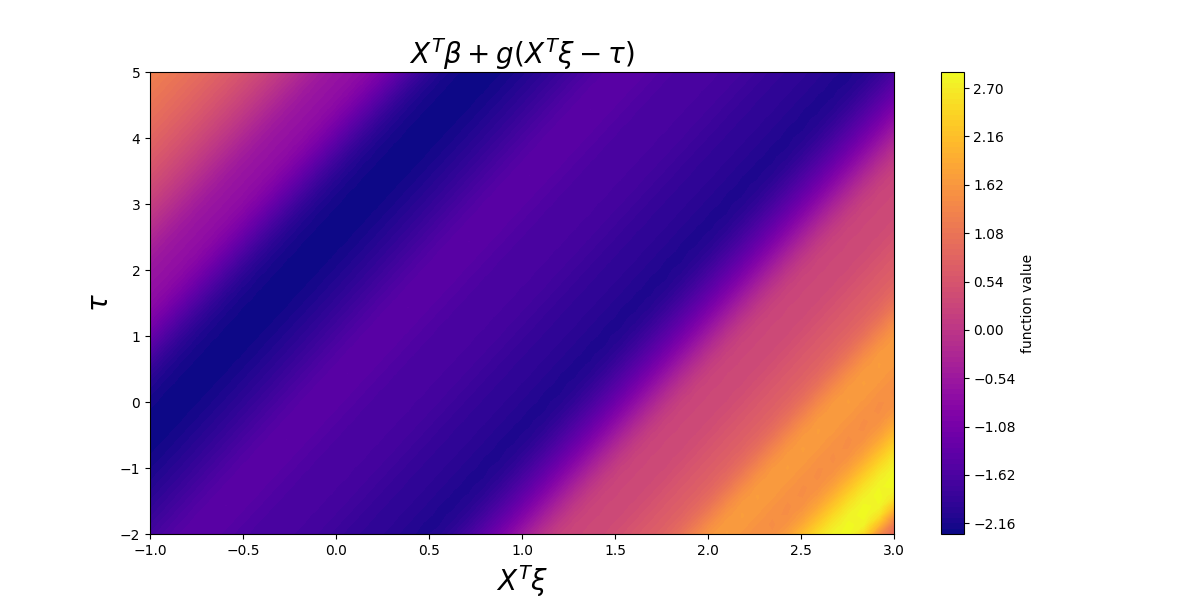}
    \caption{Estimated density function at $X^T\beta = 0.5$. The highlighted yellow region indicates improved outcomes with the corresponding treatment level $\tau$.}
    \label{fig:g_shape}
\end{figure}

For the analysis of the treatment-interaction score $X^T\xi$ in relation to the treatment $\tau$, we need to examine the function $g$ at a fixed value of $X^T\beta$ to isolate its effect. As illustrated in \figureref{fig:g_shape}, a higher warfarin dose is beneficial for patients with scores from -1.0 to 0.5. Conversely, for those with scores above 1.0, a significantly lower dosage is recommended. Therefore, in \figureref{fig:ipwc_xi_conf}, the substantial negative coefficient associated with ``age'' suggests that older patients typically require lower warfarin doses to achieve therapeutic INR levels. This observation aligns with existing literature that indicates that older patients are more sensitive to warfarin \citet{shendre2018influence} and have an increased risk of bleeding complications \citep{fihn1996risk}, necessitating lower doses for the same therapeutic effect. Likewise, the substantial negative coefficient associated with ``weight'' suggests that higher body weight typically requires higher warfarin doses to regulate their INR levels. Indeed, \citet{alshammari2020warfarin} found that obese patients required approximately 20\% higher weekly doses of warfarin compared to those with normal BMI and for each 1-point increase in BMI, the weekly warfarin dose increased by 0.69 mg \citep{mueller2014warfarin}. Furthermore, the negative coefficient for ``valve replacement'' indicates that patients with typical valve replacements might require slightly higher warfarin doses, in accordance with current guidelines that recommend targeting INR of 2.5 to 3.5 for such patients \citep{vahanian2021esc}, likely due to the increased risk of thromboembolic events associated with mechanical heart valves \citep{acc_anticoagulation_valvular_heart_disease}.

\section{Discussion and Conclusion}
This study proposes a modified partially linear model to simultaneously estimate the direct linear effects of patient covariates and a nonlinear term capturing covariate-treatment interactions. The model employs NW regression for parameter estimation and to uncover the unknown link function. Numerical simulations demonstrate empirical convergence of the estimation procedure, and the model is applied to the International Pharmacogenetics Warfarin Consortium (IPWC) dataset, focusing on Warfarin dosing decisions and their impact on achieving the target International Normalized Ratio (INR) range.

The study has several limitations. First, the parameter estimation method is primarily empirical and lacks theoretical guarantees for consistency and reliability. Nonetheless, given our numerical experiments, we believe that the estimator is indeed statistically consistent, and our future work includes theoretically studying the statistical consistency properties of our proposed approach. Second, confounder selection and feature engineering could benefit from domain-specific expertise to improve robustness. The model's clinical interpretability and practical values also require validation through prospective studies. Third, extending the framework to longitudinal data would better capture temporal dynamics in medical settings. Additionally, adapting the proposed framework to handle multi-class classification outcomes could be achieved by generalizing the logistic link function into multinomial logistic regression, thus enabling the modeling of multiple distinct clinical outcomes simultaneously. On top of that, generalizing the treatment variable to a multi-dimensional construct could enhance clinical utility in complex decision-making scenarios. Lastly, while our approach balances interpretability and efficiency, a deeper comparison with more complicated methods like deep learning models and Bayesian optimization approaches is needed to assess scalability and computational trade-offs.

\section{Acknowledgements}
This material is based upon work supported by the National Science Foundation under Grant No. DGE-2125913 and Grant No. CMMI-1847666.
\bibliography{ref}

\newpage

\startcontents[appendix]
\printcontents[appendix]{}{1}{\section*{Appendix Contents}}
\clearpage

\appendix

\section{Algorithm Overview}

We provide Algorithm \ref{alg:estimation} as an outline of the estimation method used for our semiparametric regression framework. The algorithm estimates the model parameters \(\beta\), \(\xi\) and the nonparametric function \(g\) from the data \(\{X_i, Y_i, \tau_i\}_{i=1}^n\) using kernel regression and constrained optimization with LASSO regularization.

\begin{algorithm}
\caption{Semiparametric Estimation via Constrained Optimization}\label{alg:estimation}
\begin{algorithmic}[1]
\Require Data \(\{(X_i, Y_i, \tau_i)\}_{i=1}^n\); kernel \(K(\cdot)\); bandwidth \(h\); regularization parameter \(\lambda\).
\Ensure Estimates \(\hat{\beta}\), \(\hat{\xi}\), and \(\hat{g}(\cdot)\).

\State \textbf{Compute Log-Odds:} \(\bar{Y}_i \gets \log\left(\frac{Pr(Y_i=1\mid X_i,\tau_i)}{Pr(Y_i=0\mid X_i,\tau_i)}\right)\)

\State Define the search space:
\[
\Theta = \Bigl\{ \xi \in \mathbb{R}^p : \|\xi\|_2 = 1,\; \xi_1 \ge 0 \Bigr\}.
\]

\State \textbf{Optimize \(\xi\):} Find
\[
\hat{\xi} \gets \arg\min_{\xi \in \Theta} L(\xi),
\]
where the objective function is defined as
\[
L(\xi) = \sum_{i=1}^n \left[r_i(\xi)\right]^2 + \lambda\,\|\xi\|_1,
\]
with
\[
r_i(\xi) = \hat{e}_{yi}(\xi) - \hat{e}_{xi}(\xi)^T\,\bar{\beta}(\xi).
\]
For each candidate \(\xi\), compute:
\[
Z_i(\xi) = X_i^T\xi - \tau_i,
\]
\[
\hat{e}_{xi}(\xi) = X_i - \widehat{E}\Bigl[X_i \mid Z_i(\xi)\Bigr],
\]
\[
\hat{e}_{yi}(\xi) = \bar{Y}_i - \widehat{E}\Bigl[\bar{Y}_i \mid Z_i(\xi)\Bigr],
\]
where \(\widehat{E}[\,\cdot\, \mid Z]\) is estimated via Nadaraya–Watson regression, and the least-squares estimate:
\[
\bar{\beta}(\xi) = \left(\sum_{i=1}^n \hat{e}_{xi}(\xi)\,\hat{e}_{xi}(\xi)^T\right)^{-1}\sum_{i=1}^n \hat{e}_{xi}(\xi)\,\hat{e}_{yi}(\xi).
\]

\State \textbf{Finalize \(\beta\):} Set \(\hat{\beta} \gets \bar{\beta}(\hat{\xi}) \).

\State \textbf{Estimate \(g(\cdot)\):} For any \(Z\), compute
\[
\hat{g}(Z) = \frac{\displaystyle\sum_{i=1}^n K\Bigl(\frac{Z_i(\hat{\xi})-Z}{h_i}\Bigr)\,\Bigl(\bar{Y}_i - X_i^T\hat{\beta}\Bigr)}
{\displaystyle\sum_{i=1}^n K\Bigl(\frac{Z_i(\hat{\xi})-Z}{h_i}\Bigr)},
\]
with \(Z_i(\hat{\xi}) = X_i^T\hat{\xi} - \tau_i\).

\State \Return \(\hat{\beta}\), \(\hat{\xi}\), and \(\hat{g}(\cdot)\).

\State \textbf{Inference:} Return dual score $\{X_k^T \hat{\beta}, X_k^T \hat{\xi}\}$, prediction $\text{sign}(X_k^T \hat{\beta} + \hat{g}(X_k^T \hat{\xi} - \tau_k))$, and counterfactual optimal treatment $\tilde{\tau} = \arg\max_{\tau} \hat{g}(X_k^T \hat{\xi} - \tau)$.

\end{algorithmic}
\end{algorithm}

\section{Further Discussion on Experiments}

\subsection{Experiment Settings}
In the four simulation scenarios, the number of features are 8, 8, 4, and 20, respectively. Among them, scenario 4 has 8 categorical features, while the first three scenarios only have numerical features. For each scenario, $n=\{10, 10^2, 10^3, 10^4\}$ samples are generated from the distribution specified in Table~\ref{table:table_all} as the train set. Instead of using a conventional test set, we compute the mean squared error (MSE) directly with the true function over the entire domain, defined as \( \int (\hat{g}(Z) - g(Z))^2 dZ \).

In our Warfarin dosing study, the dataset originally comprised 21 features before one–hot encoding. Among these, 4 features are continuous numerical variables: Age, Height (cm), Weight (kg), and Indication for Warfarin. There are 9 binary categorical features that serve as direct yes/no indicators: Diabetes, CHF/Cardiomyopathy, Valve Replacement, Aspirin, Amiodarone\_Cordarone, Current Smoker, Use of Statins, Use of Enzyme Inducers, and Use of Antibiotics. Finally, 8 features were originally multi-category variables that were one–hot encoded. These include Gender (e.g., male/female), Race (with categories such as Black, Unknown, and White), CYP2C9 (with its multiple allelic variants), and five separate VKORC1 genetic markers (namely -1639, 497, 1173, 1542, and 3730). The feature space comprises 30 dimensions (after one-hot encoding of categorical variables). The train set has 3177 samples, which are obtained after applying augmentation via SMOTE, and the test set has 353 samples.

There are two main hyperparameters in the proposed algorithm, a bandwidth parameter $h$ as well as a regularization parameter $\lambda$. $h$ is selected from $\{0.15, 0.2, 0.25, 0.3, 0.35, 0.4\}$ and $\lambda$ is selected from $\{1 \times 10^{-5}, 1 \times 10^{-4}, 1 \times 10^{-3}, 1 \times 10^{-2}, 1 \times 10^{-1}\}$. We used 5-fold cross-validation for hyperparameter tuning. More details on the sensitivity to these hyperparameter choices are shown in Appendix \ref{sec: hyperparameter-sensitivity}. In particular, for the TPE algorithm implemented via Hyperopt, we set \texttt{max\_evals = 200} to ensure thorough exploration of the parameter space.

All experiments were conducted locally on a MacBook Pro equipped with an Apple M3 chip with an 8-core CPU (4 high-performance cores and 4 energy-efficient cores) and 16GB of unified memory, using Python 3.9 and PyTorch.

\subsection{Simulation}

For illustration purposes, we also include a plot for scenario 3 and 4 with the learned classification decision boundary from the training data and the distribution of the test samples in \figureref{fig:sim_heat_map_dots}. We observe that all the misclassified samples are located along the decision boundary, where the samples overlap.

\begin{figure}[htbp]
\floatconts
  {fig:sim_heat_map_dots}
  {\caption{Learned classification decision boundaries for scenario 3 and scenario 4. The dashed white line indicates the decision boundary where the function value is zero, and the heatmap density represents the log-odds (higher values indicate a greater likelihood of a positive outcome).}}
  {%
    \subfigure[Scenraio 3]{\label{fig:circle}%
      \includegraphics[width=0.45\linewidth]{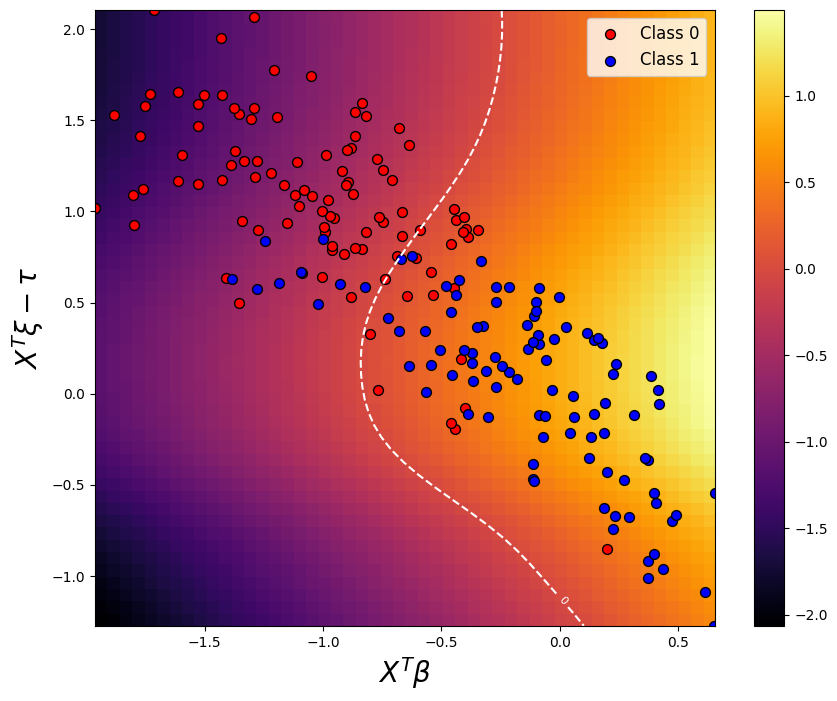}}%
    \qquad
    \subfigure[Scenraio 4]{\label{fig:square}%
      \includegraphics[width=0.45\linewidth]{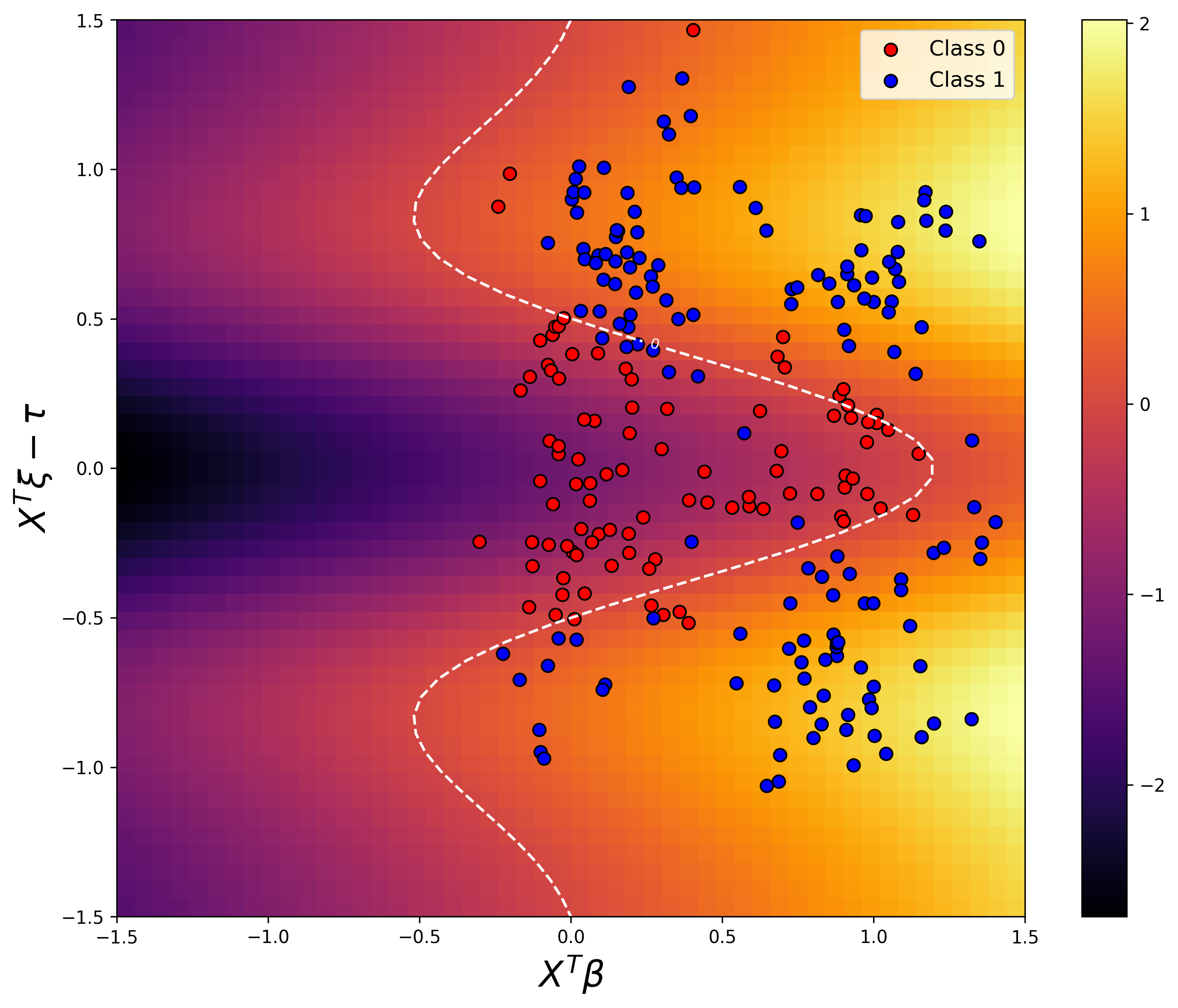}}
  }
\end{figure}

Granted, the model’s classification performance is limited by its suboptimal decision boundary. For improved distillation, knowledge transfer strategies targeting decision boundaries—such as adversarial attacks to refine boundary-critical samples—may warrant exploration \citep{heo2019knowledge}.

\subsection{Warfarin Dosing Study}

\begin{table}[t]
\centering
\footnotesize
\setlength\tabcolsep{3pt}
\renewcommand{\arraystretch}{1.1}
\caption{Performance comparison of linear models offering prognostic scores. Recall that our model is distilled from the non-interpretable expert model XGBoost, so we also include its scores here as a reference. Our model will likely improve further with a stronger expert model.}
\label{tab:perf}
\begin{tabular}{>{\raggedright\arraybackslash}p{2.2cm} *{4}{c}}
\toprule
\textbf{Model} & \textbf{Precision} & \textbf{Recall} & \textbf{F1} & \textbf{AUC} \\
\midrule
\makecell{LR}                     & 0.646  & 0.697  & 0.671  & 0.775 \\
\makecell{LR with $\tau$}      & 0.651  & \textbf{0.705}  & 0.677  & \textbf{0.788} \\
\makecell{LDA}                     & 0.671  & 0.697  & 0.684  & 0.779 \\
\makecell{LDA with $\tau$}     & 0.650  & 0.684  & 0.667  & \textbf{0.789} \\
\textbf{Our Model}                 & \textbf{0.747}  & \textbf{0.704}  & \textbf{0.725}  & \textbf{0.784} \\
\midrule
\makecell{XGBoost \\ (Reference)}  & 0.902  & 0.925  & 0.913  & 0.904 \\
\bottomrule
\end{tabular}
\end{table}

\begin{table*}[!b]
\centering
\caption{Performance Comparison of Optimization Methods for Scenario 4 (Mean $\pm$ Std)}
\label{tab:scenario4_results}
\resizebox{\textwidth}{!}{%
\begin{tabular}{lccc|ccc|ccc}
\toprule
\multirow{2}{*}{Method} &\multicolumn{3}{c|}{Runtime (s)} & \multicolumn{3}{c|}{Objective Value} & \multicolumn{3}{c}{MSE Heatmap } \\
\cmidrule(lr){2-4}\cmidrule(lr){5-7}\cmidrule(lr){8-10}
 &n=100 & n=500 & n=1000 & n=100 & n=500 & n=1000 & n=100 & n=500 & n=1000 \\
\midrule
TPE      &\textbf{0.353 $\pm$ 0.039} & \textbf{1.426 $\pm$ 0.131} & \textbf{4.405 $\pm$ 0.169} & 0.075 $\pm$ 0.047 & 0.040 $\pm$ 0.030 & 0.034 $\pm$ 0.024 & 0.149 $\pm$ 0.255 & 0.130 $\pm$ 0.106 & 0.050 $\pm$ 0.030 \\
DE       &0.373 $\pm$ 0.037 & 8.933 $\pm$ 2.226  & 34.930 $\pm$ 9.288 & \textbf{0.061 $\pm$ 0.040} & \textbf{0.021 $\pm$ 0.018} & \textbf{0.021 $\pm$ 0.026} & \textbf{0.143 $\pm$ 0.154} & \textbf{0.121 $\pm$ 0.215} & \textbf{0.032 $\pm$ 0.019} \\
CMA-ES   &0.609 $\pm$ 0.056 & 7.072 $\pm$ 0.383  & 26.004 $\pm$ 2.778 & 0.096 $\pm$ 0.076 & 0.087 $\pm$ 0.094 & 0.063 $\pm$ 0.070 & 0.299 $\pm$ 0.266 & 0.150 $\pm$ 0.112 & 0.134 $\pm$ 0.141 \\
Optuna   &0.911 $\pm$ 0.033 & 2.591 $\pm$ 0.033  & 5.861 $\pm$ 0.501 & 0.063 $\pm$ 0.040 & 0.022 $\pm$ 0.018 & 0.028 $\pm$ 0.026 & 0.145 $\pm$ 0.149 & 0.115 $\pm$ 0.194 & 0.035 $\pm$ 0.022 \\
\bottomrule
\end{tabular}%
}
\end{table*}

We first compare standard linear models—Logistic Regression (LR) and Linear Discriminant Analysis (LDA)—along with their variants that incorporate the treatment variable as a continuous covariate (denoted as ``LR/LDA with $\tau$''). In Section~\ref{sec:comp_diag_score}, we assess the clinical interpretability of their prognostic scores, where $X^T\beta$ serves as an analog to the linear predictor in logistic regression. To provide a more comprehensive evaluation, we also examine their predictive performance on classification tasks. Table~\ref{tab:perf} summarizes key metrics (precision, recall, F1, and ROC AUC), highlighting the effectiveness of our proposed model relative to these standard approaches.

\subsection{Sensitivity Analysis}
\subsubsection{Different Hyperparameters} \label{sec: hyperparameter-sensitivity}
Our proposed algorithm has two hyperparameters, a bandwidth parameter $h$ as well as a regularization parameter $\lambda$. We performed a sensitivity analysis on these two hyperparameters using the Warfarin (IWPC) dataset. The results are shown in Figure~\ref{fig:hype_sensitivity}. The top subfigure shows the average MSE over five runs, while the bottom subfigure shows the average training time. In the plots, green circles indicate lower MSE or faster training times, while red circles indicate higher MSE or longer training times. An ``X'' marks any hyperparameter combination for which the algorithm failed to converge.

Our analysis shows that our algorithm can achieve competitive MSE levels under a wide range of hyperparameters ($h \in [0.2, 0.4]$ and $\lambda \in [10^{-5}, 10^{-2}]$), demonstrating its robustness to different hyperparameter settings. Intuitively, $\lambda$ balances the sparsity of the solution and prevents overfitting, while higher $h$ offers a smoother optimization landscape at the expense of some accuracy. Additionally, all hyperparameter combinations have very similar training times. Recall that all experiments were conducted on a personal laptop, suggesting that training times could be substantially reduced using a compute server.

\begin{figure}[htbp]
\floatconts
{fig:hype_sensitivity}
  {\caption{Hyperparameter Sensitivity on IWPC.}}
  {%
    \subfigure[Mean Squared Error]{\label{fig:circle}%
      \includegraphics[width=0.9\linewidth]{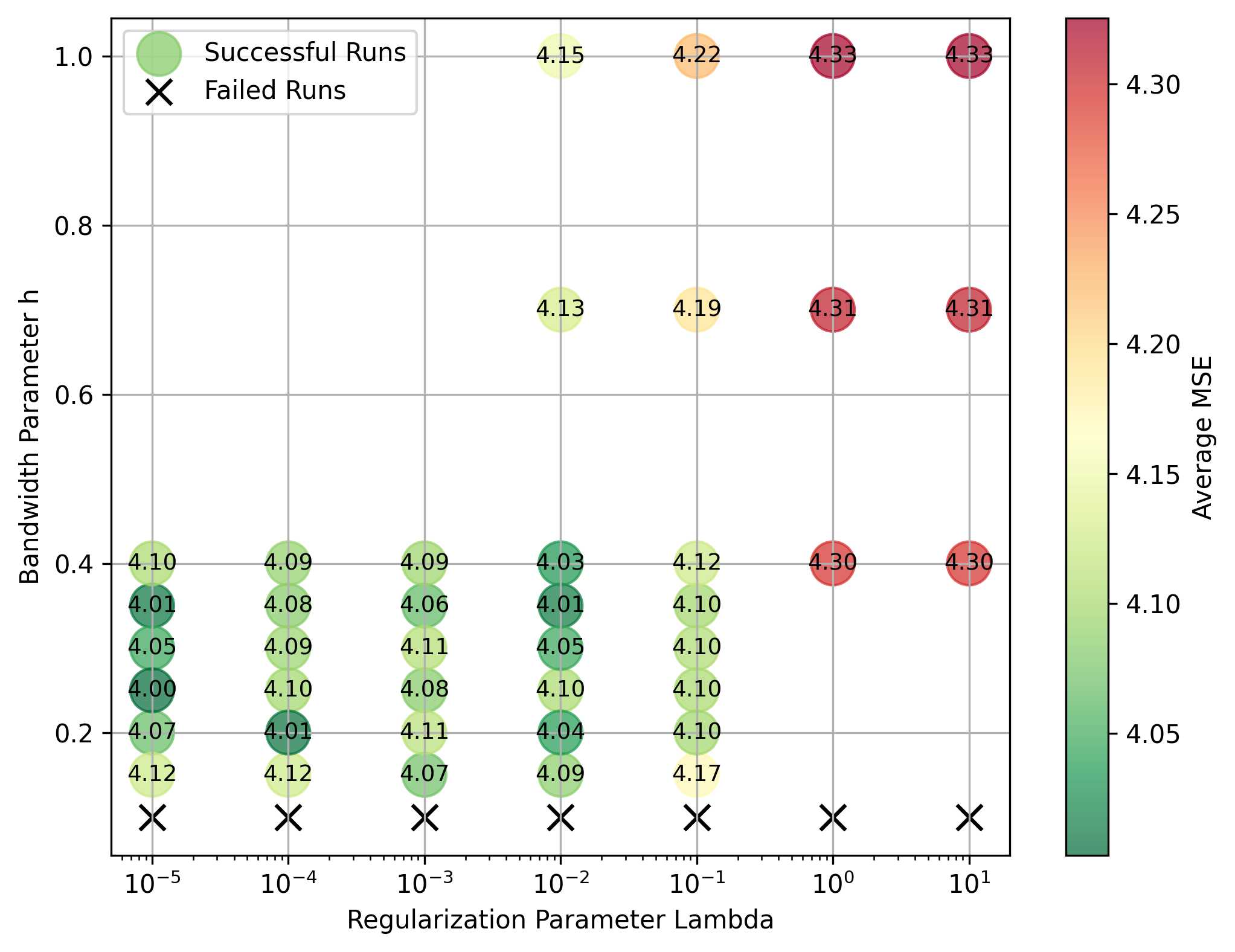}}\\[1ex]
    \subfigure[Training Time]{\label{fig:square}%
      \includegraphics[width=0.9\linewidth]{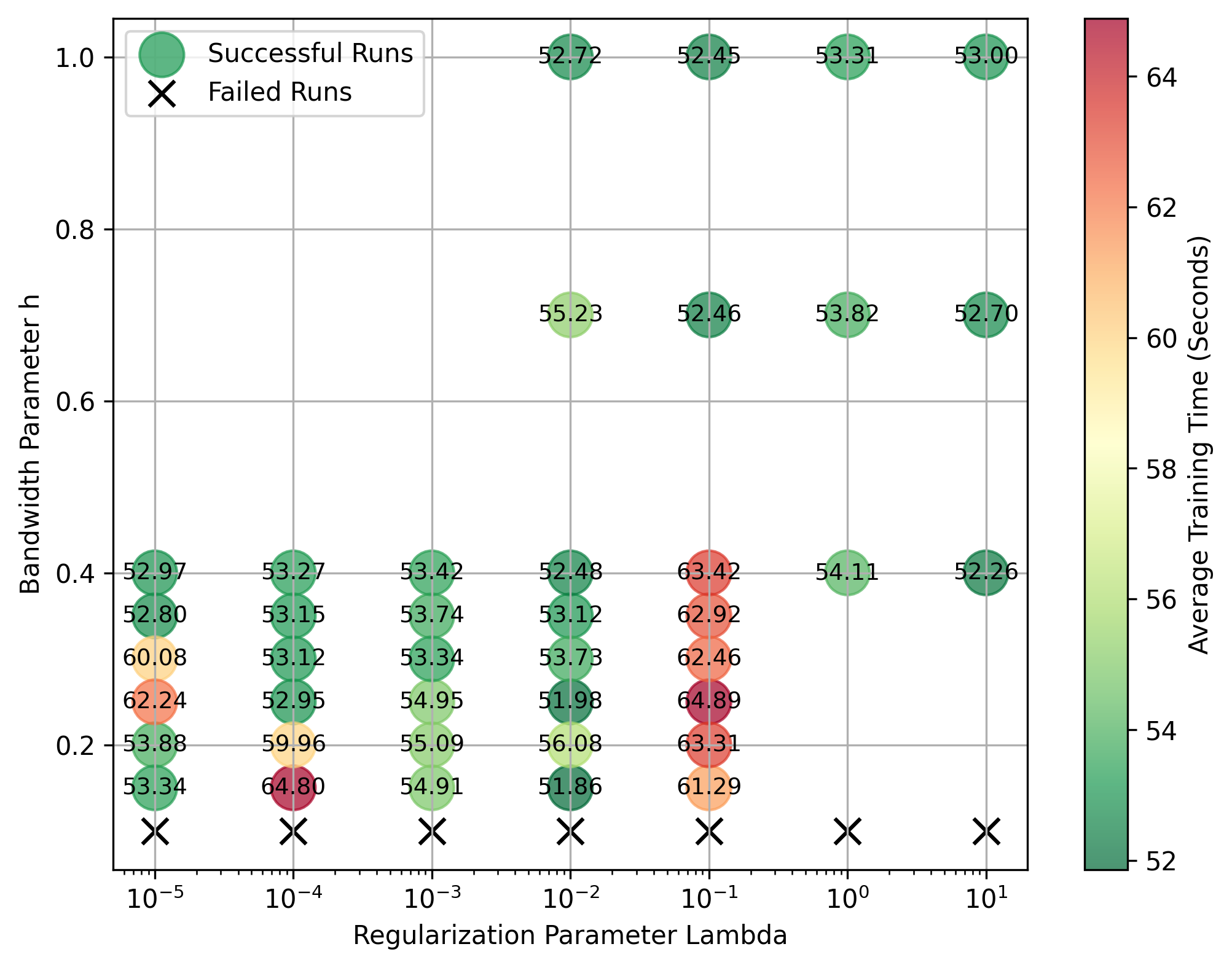}}
  }
\end{figure}

\subsubsection{Different Optimization Algorithms}

To assess the sensitivity of our constrained optimization procedure, we perform a comparative analysis of different optimization methods for solving the constrained problem in our simulation study. For Scenario 4, we run experiments using four optimization methods—Tree-based Parzen Estimators (TPE), Differential Evolution (DE), CMA Evolution Strategy (CMA-ES), and Optuna—across three sample sizes ($n=\{100, 500, 1000\}$), with five repetitions per configuration. The performance metrics reported are: total runtime, mean squared error (MSE) of the estimated heatmap versus ground truth, and the final optimal objective value (as defined in Equation \eqref{obj:constrained}). Our results in Table~\ref{tab:scenario4_results} show that TPE achieves the fastest runtime while yielding comparable MSE error (although not the best) and reasonable objective values at optimality. Notably, DE performs the best in terms of error minimization, but it is also the most time consuming; therefore, if computational time and resources are not a limiting factor, DE might be the preferred option. Conducting the same experiments in other scenarios yield similar trends.

\section{Comparison with Existing Methods}

To clarify the novelty of our approach and address concerns about its distinction from existing methods, we provide a comparative analysis of treatment-covariate interaction learning algorithms. Table~\ref{table:compare} summarizes key attributes, including their ability to handle continuous treatments, classification tasks, interpretability, and estimation strategies. Unlike prior methods, our dual-score model offers both optimal treatment rule estimation and quantitative diagnostic interpretability, making it particularly suited for clinical applications.

\begin{table*}[htbp]
\centering
\caption{Comparison of different algorithms for treatment-covariates interaction learning}
\scriptsize 
\setlength\tabcolsep{2pt} 
\renewcommand{\arraystretch}{1.1} 
\begin{tabular}{@{}cccccccc@{}}
\toprule
\makecell{Algorithm} & \makecell{Continuous\\Treatment} & \makecell{Classification\\Capability} & \makecell{Optimal Treatment\\Rule} & \makecell{Quantitative Prognostic\\Interpretability} & \makecell{Model\\Estimation} & \makecell{Data\\Type} \\
\midrule
\makecell{Clause-based Decision List \\ \citep{zhang2018interpretable}} & No & Yes & Yes & No & \makecell{Kernel Ridge Regression \\with Q-Learning} & Longitudinal \\
\makecell{AutoGluon \\\citep{erickson2020autogluon}} & Yes & Yes & No & No & Ensemble Learning & Cross-sectional \\
\makecell{Decision Trees \\ \citep{laber2015tree}} & No & Yes & Yes & No & \makecell{Tree-based Methods} & Cross-sectional\\
\makecell{SIMSL \\ \citep{park2022single}} & Yes & No & Yes & No & \makecell{Spline Regression} & Cross-sectional\\
\makecell{Change-Plane Model \\ \citep{jin2023change}} & No & Yes & No & Yes & \makecell{Spline Regression} & Longitudinal\\
\makecell{Our Model} & \textbf{Yes} & \textbf{Yes} & \textbf{Yes} & \textbf{Yes} & \makecell{NW Regression} & Cross-sectional\\
\bottomrule
\end{tabular}
\label{table:compare}
\end{table*}

\end{document}